\RequirePackage{fix-cm}
\documentclass[smallextended]{svjour3m}     
\usepackage{cite}
\usepackage{braket}
\usepackage{graphics}
\usepackage{mathptmx}
\usepackage{amsmath}
\usepackage[utf8]{inputenc}
\usepackage{xcolor}
\usepackage[bookmarks=true,bookmarksnumbered=false,bookmarksopen=false, breaklinks=false,pdfborder={0 0 0},backref=false,colorlinks=true,citecolor=red,linkcolor=blue,urlcolor=blue]{hyperref}
\usepackage[b5paper]{geometry}
\newcommand{\op}{\boldsymbol}

\newcommand{\orcid}[1]{\href{https://orcid.org/#1}{\resizebox{10px}{!}{\includegraphics{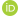}}}}
\journalname{\noindent International Journal of Theoretical Physics\\
\href{https://doi.org/10.1007/s10773-021-04906-w}
{https://doi.org/10.1007/s10773-021-04906-w} \small{(Published version available on the hyperlink)}}
\begin{document}
\title{The Delayed-Choice Quantum Eraser Leaves No Choice}
\author{Tabish Qureshi\orcid{0000-0002-8452-1078}}
\institute{Centre for Theoretical Physics, Jamia Millia Islamia, New Delhi,
India.
\email{tabish@ctp-jamia.res.in}}

\maketitle

\begin{abstract}
A realizable delayed-choice quantum eraser, using a modified
Mach-Zehnder (MZ) interferometer and polarization entangled photons, is
theoretically analyzed here. The signal photon goes through a modified MZ
interferometer, and the polarization of the idler photon provides path
information for the signal photon. The setup is very similar to the 
delayed-choice quantum eraser experimentally studied by the Vienna group.
In the class of quantum erasers with discrete output states, it is easy
to see that the delayed mode leaves no choice for the experimenter. The
which-way information is always erased, and every detected signal photon
fixes the polarization state of the idler, and thus gives information on
precisely how the signal photon traversed the two paths.
The analysis shows that the Vienna delayed-choice quantum eraser is the
first experimental demonstration of the fact that the delayed mode leaves
no choice for the experimenter, and the which-way information is always
erased. Additionally it is shown that this argument holds even in a
conventional two-slit quantum eraser. Every photon registered anywhere
on the screen, fixes the state of the two-state which-way detector
in a unique mutually unbiased basis. In the delayed-choice quantum eraser
experiments, the role of mutually unbiased basis sets for the which-way
detector, has been overlooked till now.
\end{abstract}


\section{Introduction}

The concept of wave-particle duality started out as a debate on the 
corpuscular nature versus wave nature of light. With the advent of
quantum mechanics, a new language emerged which described this concept,
namely the Bohr's principle of complementarity \cite{bohr}. According
to Bohr, the two natures of quantum objects, which we shall refer to as
quantons here, the wave and particle natures complement each other.
However, the two natures are also mutually exclusive so that an experiment
which brings out one nature, necessarily hides the other. The two-slit
interference experiment, where one additionally tries to probe which of
the two slits the quanton went through, became a test-bed for the concept of
wave-particle duality right from the time of its inception \cite{einstein}.
It soon became clear that if one tries to get the which-path or which-way
information about the quanton, the interference is destroyed.
Jaynes \cite{jaynes} came up with an interesting idea that there can be
ways in which the acquired which-way information can be erased such that
the destroyed interference can be brought back, in perfect harmony with
the concept of wave-particle duality. The modern formulation of "quantum
eraser" was proposed by Scully and Dr\"uhl \cite{druhl}. What made their
proposal more exciting was their suggestion that one may choose to erase
the which-way information much after the quanton had registered on the
screen, and the interference could still be recovered. This initiated a
lively debate on the subject which continues till date
\cite{esw,mohrhoff,srik,aharonov,hiley,ellerman,taming,kastner,kastnerbook1,kastnerbook2,TQ}.
The concept of quantum eraser implied that the experimenter could choose
to retain the which-way information or erase it, and as a consequence,
can force the quanton to behave either as a particle or a wave.
The ``delayed-choice" quantum eraser went a leap further by suggesting
that the quanton which traveled the two paths and hit the screen, may
be forced to behave like a particle or a wave by a choice made the 
experimenter after it has already hit the screen. This kind of thinking
led to a talk of "retrocausality" in the delayed-choice quantum eraser
experiment, which is still being hotly debated
\cite{ellerman,taming,kastner,kastnerbook1,kastnerbook2}.

\begin{figure}
\centerline{\resizebox{8.0cm}{!}{\includegraphics{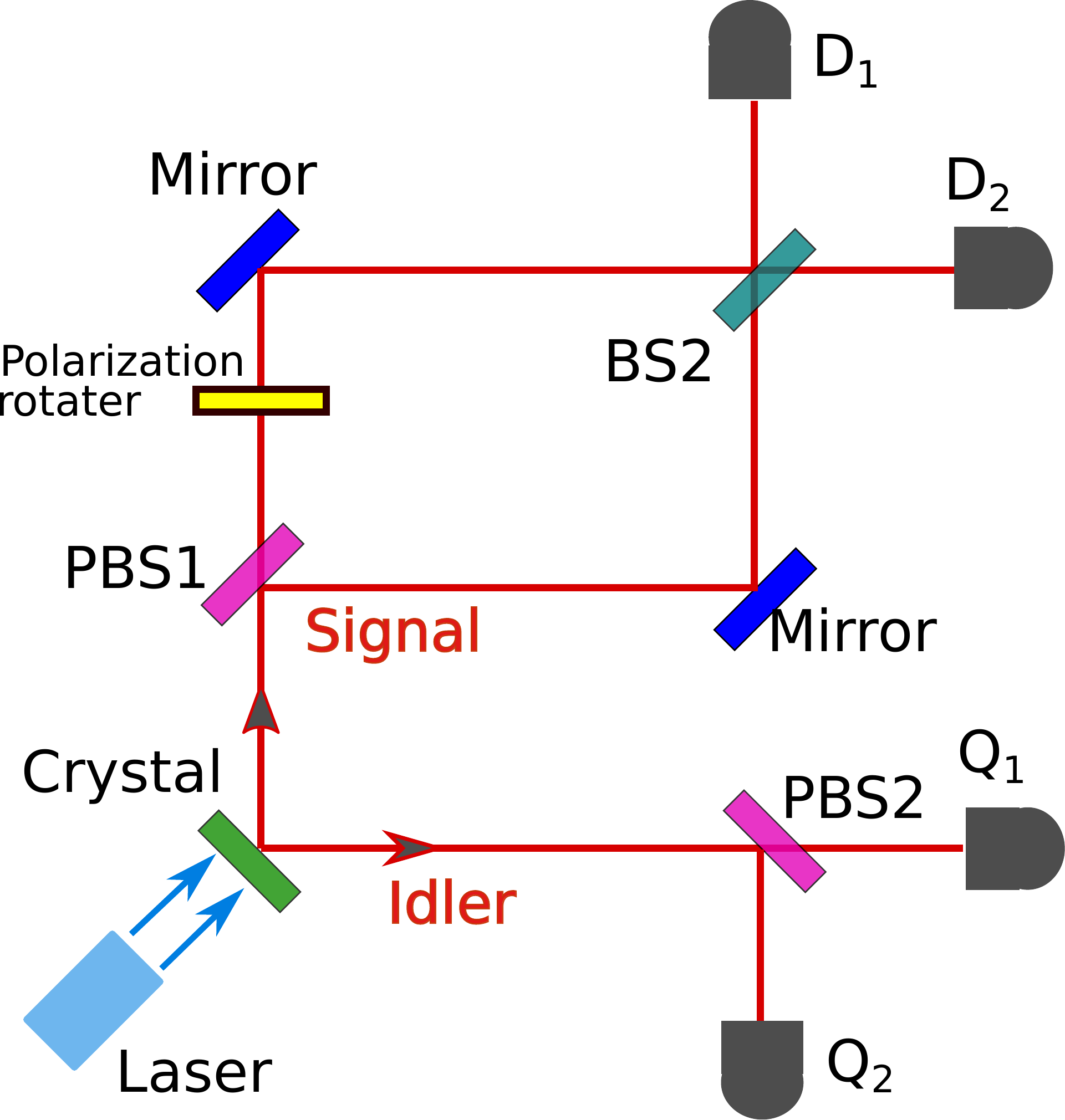}}}
\caption{A schematic diagram of a quantum eraser setup using entangled
photons, and a modified Mach-Zehnder interferometer. The two paths of the
signal photon end up getting entangled with the polarization states of
the idler photon.
 }
\label{mz}
\end{figure}

With the advances in experimental techniques the quantum eraser, with or
without delayed-choice, was realized in various ways
\cite{Ma,vienna,mandel,chiao,zeilinger,kim-shih,walborn,kim,andersen,scarcelli,neves,schneider}, and several other proposals were made 
\cite{bramon,zini,barney,chianello}. It has also been demonstrated
that the idea of quantum eraser should also work for three-path
interference \cite{3eraser}. However, no experimental progress has been
in that direction yet. It may be pertinent to mention a new class 
of delayed-choice experiments with a quantum twist, that were recently
studied \cite{terno,celeri,peruzzo,kaiser,qtwist,guo}. The idea in
those experiment was explore the possibility of a quantum superposition
of wave and particle behavior.

The current debate mainly revolves around the interpretation of
delayed-choice quantum eraser. A widely held view, due to Englert,
Scully and Walther \cite{esw}, is that the choice to retain or erase the
information regarding which of the two paths the quanton followed, always
rests with the experimenter. While this view is quite acceptable for the
normal quantum eraser, it is hard to digest for many people when applied
to the quantum eraser experiment carried out in the delayed mode. 
According to this view, even in the delayed mode, the experimenter 
chooses whether the quanton displays wave nature or particle nature.
However, the authors of this view do not comment on how one should 
interpret the "actual behavior" of the quanton in such experiments.
Questions like if the quanton shows particle nature in the delayed mode,
does it actually follow only one of the two paths, are left unanswered.
Here we re-investigate the delayed choice quantum eraser, by proposing
a realizable experiment using entangled photons, and try to find answers
to the questions which are under debate.

\section{Quantum Eraser with a Mach-Zehnder setup}

Let us consider an experimental setup as shown in Fig. \ref{mz},
where there is a spontaneous parametric down conversion (SPDC) source
producing pairs of photons which are entangled in polarization such that
the state is given by
\begin{equation}
|\Psi_0\rangle = \tfrac{1}{\sqrt{2}}(|V_s\rangle|H_i\rangle + 
|H_s\rangle|V_i\rangle
)|\psi_0\rangle|\phi_0\rangle
\label{ent0}
\end{equation}
where $H$ and $V$ denotes the horizontal and vertical polarization states,
the labels $s,i$ denote the signal and idler photons, respectively. The
spatial states for the signal and idler photons are denoted by 
$|\psi_0\rangle|\phi_0\rangle$, respectively.
A Mach-Zehnder interferometer can be easily analyzed using quantum
mechanics \cite{scarani,ferrari}. After the signal photon
passes through the polarizing beam-splitter PBS1, the state changes to
\begin{eqnarray}
|\Psi_1\rangle &=& \op{U}_{PBS1}|\psi_0\rangle \nonumber\\
 &=& \tfrac{1}{\sqrt{2}}(|H_s\rangle|V_i\rangle|\psi_1\rangle + |V_s\rangle|H_i\rangle|\psi_2\rangle)|\phi_0\rangle ,
\label{ent1}
\end{eqnarray}
where the $|\psi_1\rangle, |\psi_2\rangle$ represent the states of
the signal photon in the upper and lower path of the Mach-Zehnder 
interferometer, respectively. One would notice that the two paths of the
signal photon are now entangled with the polarization states of the two
photons. The signal photon in the upper path (path 1) passes through a
polarization rotator, possibly a half-wave plate, which rotates the
polarization by 90 degrees, flipping the 
$|H_s\rangle$ state to $|V_s\rangle$ state, so that the state now reads
\begin{eqnarray}
|\Psi_2\rangle &=& \tfrac{1}{\sqrt{2}}(|V_i\rangle|\psi_1\rangle
+|H_i\rangle|\psi_2\rangle)|V_s\rangle |\phi_0\rangle .
\label{ent2}
\end{eqnarray}
This process now entangled the two paths of the signal photon with the
polarization states of the idler photon. Interestingly, the polarization of
the signal photon is now disentangled from that of the idler. One may wonder
how one can disentangle the polarization of the signal photon from the
polarization of the idler by just a local operation on the signal
photon. It is well known that entanglement cannot be changed by a local
operation. The answer is that by this operation, the entanglement has not changed
at all, but has been \emph{transferred} from the polarization degree of
freedom to the spatial degree of freedom of the signal photon. This is
indeed possible, and similar methods have been employed before to create
\emph{hybrid entanglement} \cite{hybrid}.
Separating out
the horizontal and vertical components of the idler photon can yield
information on which of the two paths the signal photon followed, simply
because $\langle V_i|\Psi_2\rangle=|\psi_1\rangle|V_s\rangle |\phi_0\rangle$,
and $\langle H_i|\Psi_2\rangle=|\psi_2\rangle|V_s\rangle |\phi_0\rangle$.

On the other hand, (\ref{ent2}) can also be rewritten as
\begin{eqnarray}
|\Psi_2\rangle &=& \tfrac{1}{2}\{|R_i\rangle (|\psi_2\rangle - i|\psi_1\rangle)\nonumber\\
&&+ |L_i\rangle (|\psi_2\rangle + i|\psi_1\rangle)\}|V_s\rangle |\phi_0\rangle ,
\label{ent2a}
\end{eqnarray}
where $|R_i\rangle = \frac{1}{\sqrt{2}}(|H_i\rangle + i|V_i\rangle)$,
$|L_i\rangle = \frac{1}{\sqrt{2}}(|H_i\rangle - i|V_i\rangle)$
represent the left and right circular polarization states, respectively.
If one measured the circular component of polarization of the idler photon,
obtaining the state $\ket{R_i}$ would tell one that the state of the signal
photon is $\tfrac{1}{\sqrt{2}}(\ket{\psi_2}-i\ket{\psi_1})$, and
obtaining the state $\ket{L_i}$ would tell one that it is
$\tfrac{1}{\sqrt{2}}(\ket{\psi_2}+i\ket{\psi_1})$.
With this, our arrangement for obtaining path information is fully in place. 
Obtaining the state $\ket{R_i}$ of the idler photon tells us that the
signal photon followed both paths, \emph{exactly as it would, if there
were no path-detecting mechanism in place}, except for a phase difference
of $-\pi/2$ between the two paths. Obtaining the state $\ket{L_i}$
of the idler tells us again that the signal photon followed both paths, but
now with a phase difference of $\pi/2$ between the two paths.
Quantum mechanics then implies that if one measures the polarization of
the idler in the horizontal-vertical (linear polarization) basis, while
the signal photon 
is still traveling, one can force it to follow one of the two MZ paths.
On the other hand, by measuring the polarization of the idler in the
circular basis, one can force the signal photon to follow both the paths.
The choice lies with the experimenter.

The effect of the second beam-splitter, on the two components
$|\psi_1\rangle, |\psi_2\rangle$, is the following \cite{scarani}
\begin{eqnarray}
\op{U}_{BS2}|\psi_1\rangle &=& \tfrac{1}{\sqrt{2}}(|D_1\rangle + i|D_2\rangle) \nonumber\\
\op{U}_{BS2}|\psi_2\rangle &=& \tfrac{1}{\sqrt{2}}(i|D_1\rangle + |D_2\rangle) ,
\label{ubs2}
\end{eqnarray}
where $\op{U}_{BS2}$ represents the unitary evolution due to the mirrors
and the second beam-splitter BS2, and $|D_1\rangle, |D_2\rangle$ are the
states at the detectors $D_1, D_2$, respectively. If the state of the
signal photon is $|\psi_1\rangle$ or $|\psi_2\rangle$, in both the situations
it is equally likely to hit $D_1$ or $D_2$. This implies a loss of interference,
resulting from extraction of which-path information by the idler.
So, obtaining the horizontal or vertical state of the idler, destroys the
interference of the signal photon, but yields its precise path information.

However, if one measured the circular polarization of the idler, and 
obtained  the state $\ket{R_i}$, it would tell us that the state of the
signal photon would be $\tfrac{1}{\sqrt{2}}(\ket{\psi_2}-i\ket{\psi_1})$.
The second beam-splitter BS2 would take this state only to $D_2$: 
$\op{U}_{BS2}\tfrac{1}{\sqrt{2}}(\ket{\psi_2}-i\ket{\psi_1})=\ket{D_2}$.
This implies interference with the bright fringe at $D_2$ and the dark one at $D_1$.
On the other hand, obtaining the state $\ket{L_i}$ would tell us that the
state of the
signal photon would be $\tfrac{1}{\sqrt{2}}(\ket{\psi_2}+i\ket{\psi_1})$. The
beam-splitter BS2 would take this state only to $D_1$: 
$\op{U}_{BS2}\tfrac{1}{\sqrt{2}}(\ket{\psi_2}+i\ket{\psi_1})=i\ket{D_1}$.
This also implies interference, but with the bright fringe at $D_1$ and the
dark one at $D_2$.
Both these situations describe the phenomenon of quantum erasure, where the
lost interference comes back if the which-path information is erased.
However, the two interferences are mirror images of each other, and taken
together, they cancel each other out.

Let us now look at the delayed mode where no measurement is made on the 
idler photon, the path of the idler being much longer, and the signal photon
reached the detectors. For example, in one experiment performed by the
Vienna group \cite{vienna}, the idler photon travels a distance of 144
kilometers before it reaches the analyzing detectors, whereas the MZ paths
are of the order of 2 meters. In our setup, the final state of the two
photons, just before the signal photon hits the detectors, is given by
\begin{eqnarray}
|\Psi_3\rangle &=& \op{U}_{BS2}|\Psi_2\rangle = \tfrac{1}{\sqrt{2}}
\op{U}_{BS2}(|V_i\rangle|\psi_1\rangle
+ |H_i\rangle|\psi_2\rangle)|V_s\rangle |\phi_0\rangle \nonumber\\
 &=& \tfrac{1}{2}[|V_i\rangle(|D_1\rangle+i|D_2\rangle)
+|H_i\rangle(i|D_1\rangle+|D_2\rangle)]|V_s\rangle |\phi_0\rangle. 
\label{ent3}
\end{eqnarray}
This state indicates that $D_1$ and $D_2$ are equally likely to register the
signal photon, as
$|\langle D_1|\Psi_3\rangle|^2 = |\langle D_2|\Psi_3\rangle|^2 =1/2$,
which in turn implies no interference.

An interesting scenario emerges if one rewrites the state (\ref{ent3}) as
\begin{eqnarray}
|\Psi_3\rangle &=& \tfrac{1}{\sqrt{2}}\left(i|D_1\rangle|L_i\rangle +
|D_2\rangle |R_i\rangle\right)|V_s\rangle |\phi_0\rangle. 
\label{ent4}
\end{eqnarray}
This state indicates that if the signal photon registers at $D_2$, it fixes the
polarization state of the idler to the right circular state $|R_i\rangle$,
and if it registers at $D_1$, it fixes polarization state of the idler to
the left circular state $|L_i\rangle$. But the states
$|R_i\rangle,|L_i\rangle$ correspond to the erased which-path information,
and tell us that the signal photon followed both the paths, and not one
of the two.
In the delayed mode, the experimenter no longer has the choice to seek
either which-path information or quantum eraser. This runs counter to
the widely accepted notion that the choice of which-path information or
quantum eraser, lies with the experimenter in the delayed mode \cite{esw}.
Not only does the registered signal photon tell us the that the which-path
information is erased, it tells us precisely how the signal photon traversed
the two paths, and the phase difference between the two paths, by
virtue of (\ref{ent2a}). This correlation, of course, can also be used
to recover the lost interference, constituting the usual quantum eraser.

\begin{figure}
\centerline{\resizebox{9.0cm}{!}{\includegraphics{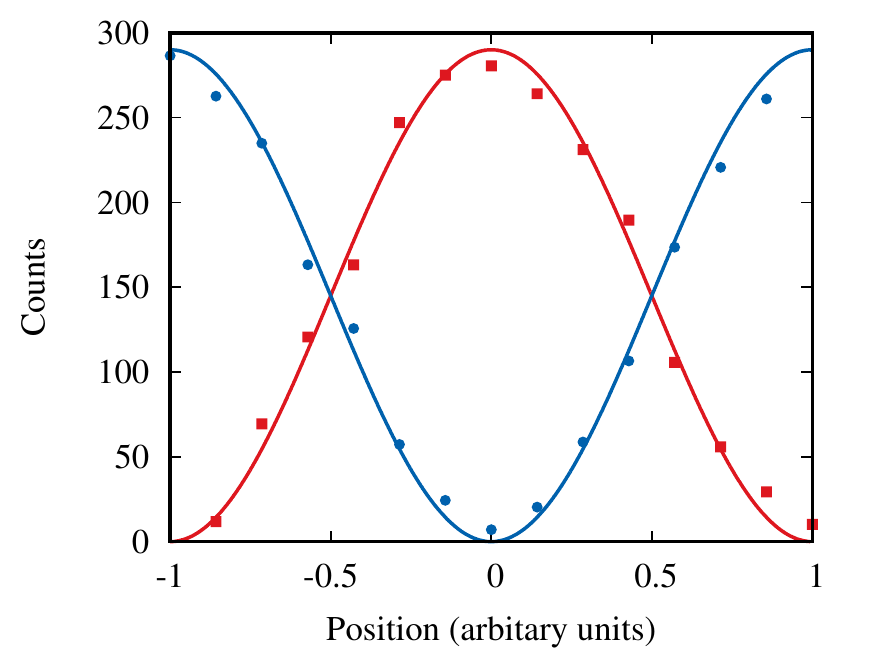}}}
\caption{Simulated results of the quantum eraser setup using entangled
photons, and a modified Mach-Zehnder interferometer (see FIG. \ref{mz}).
Points represent
coincident counts of $D_1$ (red squares) and $D_2$ (blue circles) with
one particular state of the idler, as a function of the position of BS1.
Solid lines are the corresponding 
theoretical curves. Compare with figure 3D of Ref. \cite{vienna}. The
peak will be shifted if the coincidence is done with a state of a
different mutually unbiased basis of the polarization of the idler photon.
 }
\label{vienna}
\end{figure}

More interesting is the fact that the correlation between the left-right
circularly polarized states of the idler, and the detectors $D_1$ and $D_2$
of the modified MZ setup has already been experimentally observed in
the Vienna delayed choice quantum eraser experiment \cite{vienna}.
However, its implication was not recognized for want of an analysis
similar to the one presented here. The equivalence between the setup
studied here and the one implemented by the Vienna group can be easily
seen. They used, what they call, a 'hybrid entangler' to achieve entanglement
between the two paths of one photon and the polarization states of a
causally disconnected photon. Although the authors go an extra step by
varying the position of PBS1, and observing the counts of
each detector (coincident with remote photon), there is a central position
of PBS1 for which one detector gives maximum counts, and the other one
gives almost zero (see figure 3D of Ref. \cite{vienna}). FIG. \ref{vienna}
shows simulated data of our suggested experiment, and is closely similar
to the Vienna experiment results. The maximum and minimum counts at the
central position
are the equivalent of the bright and dark fringe
of the traditional two-slit experiment. At this position of PBS1, registering
of a photon at a particular detector, fixes the polarization state of 
the other photon which is 144 km away. In the experiment, this emerges
as the prefect correlation between the two observations.

The prevalent belief \cite{esw} says that even in the delayed mode, 
observing the idler in the horizontal-vertical basis, gives one the
path-information about the signal photon. The preceding analysis shows
that this is incorrect. For example, if the signal photon registers at
$D_2$, (\ref{ent4}) tells us that the polarization state of the idler is
$|R_i\rangle$. Since
$|R_i\rangle=\frac{1}{\sqrt{2}}(|H_i\rangle+i|V_i\rangle)$, if one insists
on measuring the polarization in the horizontal-vertical basis, one
will get the two results with equal probability, and hence no interference
to speak of. However, in this case, getting a $|H_i\rangle$ or $|V_i\rangle$
does not give one any path information. This is simply because there is
a  correlation between $|D_1\rangle, |D_2\rangle$ and
$|R_i\rangle, |L_i\rangle$ (by virtue of (\ref{ent4})), and getting a (say) $|D_1\rangle$ destroys
the possibility of using (\ref{ent2}) to infer path information \cite{TQ}.
So, the loss of interference is not because of obtaining any path information
by looking at $|H_i\rangle$ and $|V_i\rangle$ states. The interference is
lost anyway, unless one correlates with the $|R_i\rangle,|L_i\rangle$ states.

\section{Movable beam-splitter}

An objection can be raised that the preceding analysis holds only for certain
fixed locations of BS1 or BS2, as only for those locations one of the detectors
$D_1, D_2$ will show zero count (destructive interference). In the following
we will show that that is not the case, and this argument can be made quite
general. Let us suppose that the beam-splitter BS1 is movable, so that its
position leads to a phase factor of $e^{2\pi ix/\lambda}$ for the upper path,
where $\lambda$ is the wavelength of the light used in the experiment.
For $x=0$ the two path lengths are the same, and all the preceding
arguments go through. For an arbitrary $x$, the final state of the two
photons, just before the signal photon hits the detectors, instead of
(\ref{ent3}), is now given by
\begin{eqnarray}
|\Psi_3'\rangle &=& \tfrac{1}{\sqrt{2}}
\op{U}_{BS2}(|H_i\rangle|\psi_2\rangle e^{\frac{2\pi ix}{\lambda}}
+ |V_i\rangle|\psi_1\rangle)|V_s\rangle |\phi_0\rangle \nonumber\\
 &=& \tfrac{1}{2}[|H_i\rangle e^{\frac{2\pi ix}{\lambda}}(i|D_1\rangle+|D_2\rangle) 
+|V_i\rangle(|D_1\rangle+i|D_2\rangle)]|V_s\rangle |\phi_0\rangle. 
\label{ent3p}
\end{eqnarray}
In terms of the states $|R_i\rangle, |L_i\rangle$, the above can be
written as
\begin{eqnarray}
|\Psi_3'\rangle &=& \tfrac{1}{2} \Big[i
|D_1\rangle\big\{(e^{\frac{2\pi ix}{\lambda}}-1)|R_i\rangle + 
(e^{\frac{2\pi ix}{\lambda}}+1)|L_i\rangle \big\} \nonumber\\
&&+ |D_2\rangle \big\{(e^{\frac{2\pi ix}{\lambda}}+1)|R_i\rangle + 
(e^{\frac{2\pi ix}{\lambda}}-1)|L_i\rangle \big\}
	\Big]|V_s\rangle |\phi_0\rangle .
\label{ent4p}
\end{eqnarray}
Now there is no correlation between the states $|R_i\rangle, |L_i\rangle$ and
the detector states $|D_1\rangle, |D_2\rangle$. So a signal photon registered
at the detectors cannot not tell us if the state of the idler will be
$|R_i\rangle$ or $|L_i\rangle$.

However, one should realize that there is nothing sacred about the basis
$|R_i\rangle, |L_i\rangle$ chosen for the idler photon. Given the polarization
states $|H_i\rangle, |V_i\rangle$, there exist an infinite number of 
\emph{mutually unbiased} basis states which can be used for the purpose.
The basis defined by $|R_i\rangle, |L_i\rangle$ happens to be just one such
basis which is unbiased with respect to $|H_i\rangle, |V_i\rangle$.
One might as well choose the following basis states for the polarization
of the idler photon
\begin{eqnarray}
|P_i\rangle &=& \frac{1}{\sqrt{2}}(e^{i\theta}|H_i\rangle + i|V_i\rangle),\nonumber\\
|Q_i\rangle &=& \frac{1}{\sqrt{2}}(e^{i\theta}|H_i\rangle - i|V_i\rangle),
\label{PQ}
\end{eqnarray}
where $\theta$ is an arbitrary phase factor.
The state of the two photons, just before the signal photon enters BS2
\begin{eqnarray}
|\Psi_2'\rangle &=& \tfrac{1}{\sqrt{2}}(e^{\frac{2\pi ix}{\lambda}}|H_i\rangle|\psi_2\rangle
+|V_i\rangle|\psi_1\rangle)|V_s\rangle |\phi_0\rangle ,
\label{ent2p}
\end{eqnarray}
can be written in terms of this new basis as
\begin{eqnarray}
|\Psi_2'\rangle &=& \tfrac{1}{2}\{|P_i\rangle (e^{i(\frac{2\pi x}{\lambda}-\theta)} |\psi_2\rangle - i|\psi_1\rangle)\nonumber\\
&&+ |Q_i\rangle (e^{i(\frac{2\pi x}{\lambda}-\theta)} |\psi_2\rangle + i|\psi_1\rangle)\}|V_s\rangle |\phi_0\rangle .
\label{ent2aq}
\end{eqnarray}
Now, if one chooses the basis such that $\theta = \frac{2\pi x}{\lambda}$,
the state of the two photons, before the signal photon enters BS2, is given by
\begin{eqnarray}
|\Psi_2'\rangle &=& \tfrac{1}{2}\{|P_i\rangle (|\psi_2\rangle - i|\psi_1\rangle)
+ |Q_i\rangle (|\psi_2\rangle + i|\psi_1\rangle)\}|V_s\rangle |\phi_0\rangle .
\label{ent2aqp}
\end{eqnarray}
It shows that for every position of the beam-splitter BS1, there exists a
basis (\ref{PQ}) for the idler, the states of which 
get correlated to $\tfrac{1}{\sqrt{2}}(|\psi_2\rangle + i|\psi_1\rangle)$ and
$\tfrac{1}{\sqrt{2}}(|\psi_2\rangle - i|\psi_1\rangle)$, exactly as
$|R_i\rangle,|L_i\rangle$ did in (\ref{ent2a}),
and hence are indicators of the signal photon following both paths.
After the signal photon passes through BS2 and is about to hit the detectors
$D_1, D_2$, the combined state of the two is given by
\begin{eqnarray}
|\Psi_3'\rangle &=& \tfrac{1}{\sqrt{2}}\left(i|D_1\rangle|Q_i\rangle +
|D_2\rangle |P_i\rangle\right)|V_s\rangle |\phi_0\rangle. 
\label{ent4q}
\end{eqnarray}
This means that every signal photon registered at the detectors, fixes
the polarization state of the idler in this particular basis. So, for
every position of the beam-splitter BS1, there exists a basis (\ref{PQ})
for the idler, the states of which are perfectly correlated with two
detectors of the signal photon. While it is true that (\ref{ent4p})
also indicates that every signal photon registered at (say) $D_1$,
fixes the state of the idler, but there is no way to know, a priori,
how that state would be read. In the present case, looking at the
position of the beam-splitter BS1, one can choose the basis (\ref{PQ})
in which to measure the polarization of the idler so that the results 
of the two are perfectly correlated. Choosing the basis may amount to
choosing the angle by which the polarization of the idler has to be rotated,
or something equally straightforward. Each detected signal photon tells
one whether the state of the idler is $|P_i\rangle$ or $|Q_i\rangle$,
and consequently also tells that the signal photon followed both paths,
and not one of the two. Since this experiment can be
performed, and the correlation measured, at least in principle, it
tells us that in the delayed mode of the quantum eraser, which-way
information is \emph{always} erased.

\section{The two-slit which-way experiment}

\begin{figure}
\centerline{\resizebox{6.0cm}{!}{\includegraphics{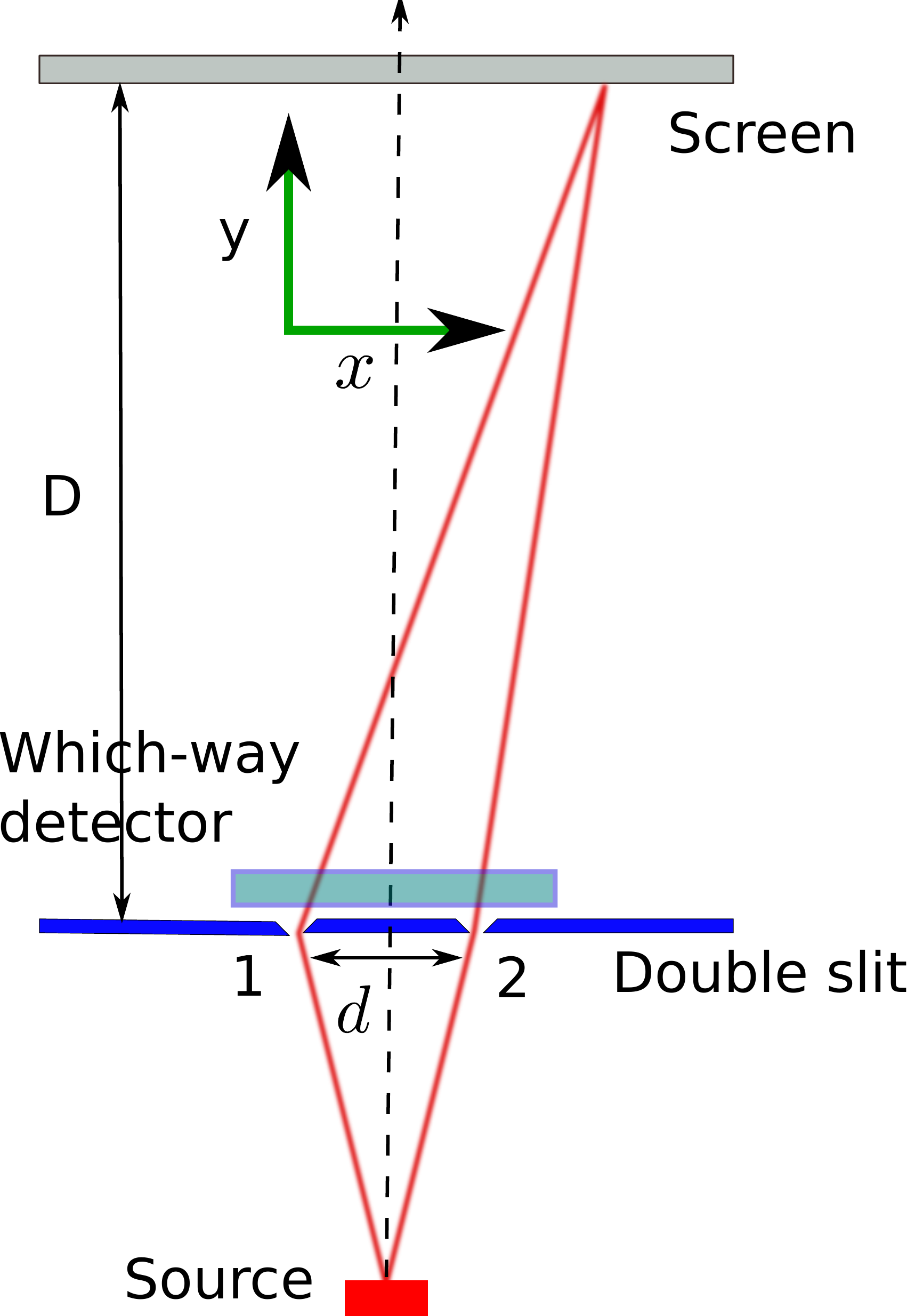}}}
\caption{A schematic diagram of a two-slit interference experiment with
which-way detection.  The two paths of the photon get entangled with the
two states of the which-way detector.
 }
\label{twoslit}
\end{figure}

\begin{figure}
\centerline{\resizebox{9.0cm}{!}{\includegraphics{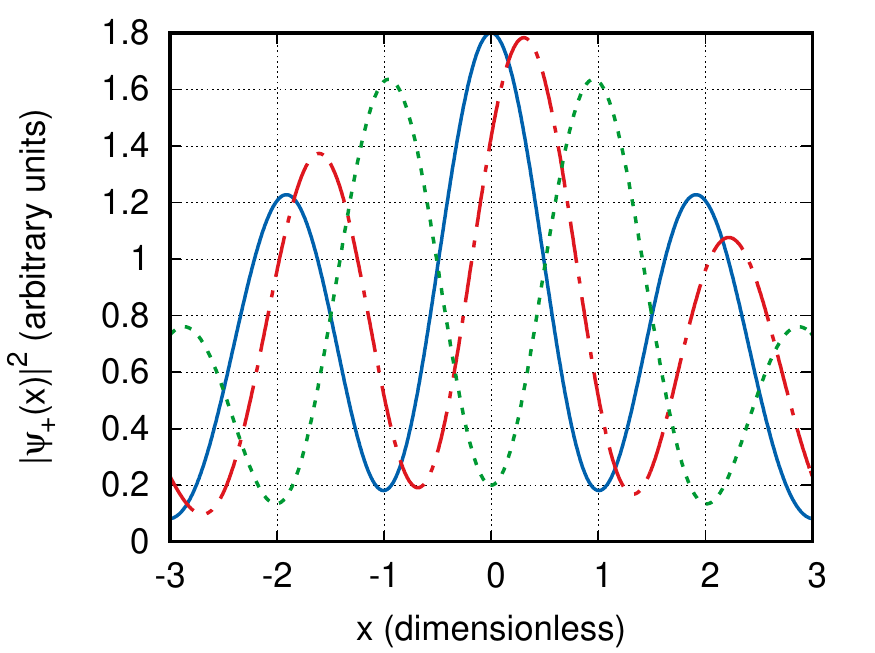}}}
\caption{Typical recovered two-slit interference because of quantum
erasure. Recovered interference $|\psi_+(x)|^2$ for $\theta=0$ 
(solid blue line), $|\psi_-(x)|^2$ for $\theta=0$ (dotted green line),
and $|\psi_+(x)|^2$ for $\theta\neq 0$ (dot-dashed red line).
 }
\label{interf}
\end{figure}

Various other delayed choice quantum eraser experiments have been performed
using conventional double-slit interference. One might wonder if the
arguments presented in the preceding discussion hold for the two-slit
delayed choice quantum eraser experiments too. This is the question we
address in the following analysis. Consider a two-slit interference
experiment with a two-state which-way detector, as shown in FIG.
\ref{twoslit}. Without specifying the nature of the which-way detector,
we assume that its effect is to entangle the two photon paths with the
two states of the which-way detector, such that the combined state, when
the photon reaches the screen, is given by
\begin{equation}
\Psi(x) = \tfrac{1}{\sqrt{2}}( \psi_1(x)|d_1\rangle + \psi_2(x)|d_2\rangle),
\label{2state1}
\end{equation}
where $|d_1\rangle, |d_2\rangle$ are orthonormal states of the which-way
detector. One can now define a mutually unbiased basis by the states
$|d^{\theta}_{\pm}\rangle = \frac{1}{\sqrt{2}}(e^{i\theta}|d_1\rangle \pm |d_2\rangle)$. The state
of the photon and which-way detector may be rewritten in the new basis
\begin{eqnarray}
\Psi(x) &=& \tfrac{1}{2}\Big([e^{-i\theta}\psi_1(x)+\psi_2(x)]|d^{\theta}_+\rangle
+ [e^{-i\theta}\psi_1(x)-\psi_2(x)]|d^{\theta}_-\rangle\Big).
\label{2state2}
\end{eqnarray}
The two states of the photon, corresponding to the which-way detector
states $|d_{\pm}^{\theta}\rangle$ are given by
$\psi_{\pm}(x) = \tfrac{1}{\sqrt{2}}[e^{-i\theta}\psi_1(x) \pm \psi_2(x)]$,
respectively. One can do a rigorous wave-packet analysis of the dynamics
of the photon, and find the two states to have the following typical
form \cite{einstein}
\begin{eqnarray}
\psi_{\pm}(x) = A(x)[1 \pm \cos\left(\tfrac{2\pi xd}{\lambda D} - \theta\right)],
\end{eqnarray}
where $d$ is the separation between the two slits, $D$ is the distance
between the slits and the screen, and $A(x)$ is an envelope function.
For $\theta=0$, $\psi_+(x)$ represents an interference pattern with a
central peak at $x=0$. On the other hand, $\psi_-(x)$ represents a similar,
but shifted interference pattern, with a \emph{minimum} at $x=0$ 
(see FIG. \ref{interf}).
One can see that if a photon is detected at $x=0$, it can only belong to
$\psi_+(x)$, because $|\psi_-(x=0)|^2 = 0$. Then, from (\ref{2state2})
one infers that the state of the which-way detector is $|d^0_+\rangle$,
and not $|d^0_-\rangle$. One can then conclude that the photon traveled
both the paths, and not one of the two. The same argument can be
made for all values of $x$ where $\psi_+(x)$ has a peak.

But what about the photons which land at positions where $\psi_+(x)$
does \emph{not} have a peak? In that case one can choose a different
basis for the which-way detector states such that
$\theta = \tfrac{2\pi xd}{\lambda D}$. Remember that the interference patterns
are obtained \emph{only in coincidence} with the which-way detector states,
and in coincidence with $|d^{\theta}_{\pm}\rangle$, the interference patterns
will be shifted (see FIG. \ref{interf}). They will be shifted in such a way
that $|\psi_-(x)|^2 = 0$ for that particular $x$. One can then logically
conclude that the state of the which-way detector is $|d^{\theta}_{+}\rangle$,
and not $|d^{\theta}_{-}\rangle$, and the which-way information is erased.
This again indicates that the photon followed both the paths. This
correlation can be experimentally seen, as looking at the values of $x$
at which the photon is registered, one can choose a mutually unbiased
basis of the which-way detector states $|d^{\theta}_{\pm}\rangle$,
specified by $\theta = \tfrac{2\pi xd}{\lambda D}$. Thus, for every
photon registered on the screen, one can choose a mutually unbiased
basis of the which-way detector states, whose measurement result is
predicted by the registered photon. So, even in a two-slit delayed-choice
quantum eraser, every registered photon fixes the state of the which-way
detector in a knowable basis, and thus always erases the which-way
information.

Some comments on how one can choose a different mutually unbiased basis
for the which-way detector in delayed-choice quantum eraser experiments.
In the experiment of Kim et.al. \cite{kim-shih}, the idler photon, after
traversing two paths, is recombined using a beam-splitter. Changing the
relative lengths of the two paths would amount to choosing a different
mutually unbiased basis for the idler. The recombining beam-splitter may
be moved in synchrony with the movable detector for the signal photon
\cite{kim-shih}.
In the experiment of Scarcelli et.al. \cite{scarcelli}, quantum erasing is
achieved by letting the idler pass through a fixed narrow slit, and observing
the signal photon in coincidence with it. Here, changing the basis of the
path-detecting photon (idler) can be achieved by changing the position of
the narrow slit through which the idler passes. 

\section{Conclusions}

The correlation in (\ref{ent4}) emerges in a straightforward fashion
in a MZ like setup where there are only two discrete output states.
The confusion prevailing in the literature regarding the delayed choice
quantum eraser experiments, can be resolved by recognizing that the
quantum eraser in a MZ like setup may be summarized by writing the
state (\ref{ent4}) in two complementary bases,
$i|D_1\rangle|L_i\rangle + |D_2\rangle |R_i\rangle = 
|\psi_1'\rangle|V_i\rangle + |\psi_2'\rangle|H_i\rangle$, (ignoring
normalization) where $|\psi_1'\rangle,|\psi_2'\rangle$ represent a
complementary basis for the signal, and are proxies for the two paths. 
The idler is prepared in a basis depending on the basis in which the
signal is measured, and vice-versa.
Thus the states $|V_i\rangle,|H_i\rangle$ of the idler, provide path
information of the signal photon, and
$|L_i\rangle,|R_i\rangle$ represent erased path information. This makes it
easy to see that the delayed mode, when the signal photon ends up
in $|D_1\rangle$ or $|D_2\rangle$, can only result in erased path information. 
There is no choice left.
As the Vienna experiment \cite{vienna} is the first one to implement
a delayed-choice
quantum eraser in a MZ setup, it is also the first one to experimentally 
demonstrate that in the delayed mode, which-way information is always
erased, and the photon always follows both the paths. 
In the traditional two-slit experiment, this effect is hidden because
there is a continuous set of positions on the screen where the photon
can register. However, we have shown that even in the two-slit delayed
choice quantum eraser, for every photon detected on the screen, there
exists a basis in which the state of the which-way detector gets fixed
by the act of photon hitting the screen. This basis can be known from
the position of the photon, and the corresponding measurement can be
made on the which-way detector to test the correlation.
In the light of
this analysis, and the results of the Vienna experiment, the long held
notion that in the delayed mode, the experimenter has a choice between
reading the which-way information or erasing it, should be given up. 
In the delayed mode, the which-way information is always erased. This
takes the mystery out of the delayed-choice quantum eraser, and renders
irrelevant any talk of retrocausality.








\begin{thebibliography}{0}

\bibitem{bohr} N. Bohr, ``The quantum postulate and the recent development of
atomic theory,"
\href{https://doi.org/10.1038/121580a0}
{\emph{Nature (London)} \textbf{121}, 580-591 (1928). }

\bibitem{einstein} T. Qureshi, R. Vathsan, ``Einstein's recoiling slit experiment, complementarity and uncertainty," \href{https://www.doi.org/10.12743/quanta.v2i1.11}{\emph{Quanta} \textbf{2}, 58-65 (2013)}. 

\bibitem{jaynes} E. Jaynes, in
{\emph{Foundations of Radiation Theory and
Quantum Electrodynamics}, ed. A.O. Barut (Plenum, New York 1980), pp. 37.}

\bibitem{druhl} M. O. Scully and K. Dr\"{u}hl, ``Quantum eraser: A proposed photon correlation experiment concerning observation and "delayed choice" in quantum mechanics,"
\href{https://doi.org/10.1103/PhysRevA.25.2208}
{\emph{Phys. Rev. A} \textbf{25}, 2208 (1982).}


\bibitem{esw} B.-G. Englert, M. O. Scully, H. Walther, ``Quantum erasure in double-slit interferometers with which-way detectors,"
\href{https://doi.org/10.1119/1.19257}
{\emph{Am. J. Phys.} \textbf{67}, 325 (1999).}

\bibitem{mohrhoff} U. Mohrhoff, ``Objectivity, retrocausation, and the experiment of Englert, Scully, and Walther,"
\href{https://doi.org/10.1119/1.19258}
{\emph{Am. J. Phys.} \textbf{67}, 330 (1999).}

\bibitem{srik} R. Srikanth, ``A quantum field theoretic description of the delayed choice experiment,"
{\emph{Curr. Sci.} \textbf{81}, 1295 (2001).}

\bibitem{aharonov} Y. Aharonov, M.S. Zubairy,
``Time and the quantum: erasing the past and impacting the future,"
\href{https://doi.org/10.1126/science.1107787}
{\emph{Science}, 307(5711):875 (2005).}

\bibitem{hiley} B.J. Hiley, R.E. Callaghan, 
``What is erased in the quantum erasure?,"
\href{https://doi.org/10.1007/s10701-006-9086-4}
{\emph{Found. Phys.}, \textbf{36(12)} 1869 (2006).}

\bibitem{ellerman} D. Ellerman, ``Why delayed choice experiments do Not imply retrocausality,"
\href{https://doi.org/10.1007/s40509-014-0026-2}
{\emph{Quantum Stud.: Math. Found.} \textbf{2}, 183 (2015).}

\bibitem{taming} J. Fankhauser, ``Taming the delayed choice quantum eraser,"
\href{https://doi.org/10.12743/quanta.v8i1.88}
{\emph{Quanta} \textbf{8}, 44 (2019)}. 

\bibitem{kastner} R.E. Kastner, ``The ‘delayed choice quantum eraser’ neither erases nor delays,"
\href{https://doi.org/10.1007/s10701-019-00278-8}
{\emph{Found. Phys.} \textbf{49}, 717 (2019).}

\bibitem{kastnerbook1} R.E. Kastner,
{\emph{The Transactional Interpretation of Quantum Mechanics: The Reality of Possibility} (Cambridge University Press, Cambridge, 2012).}

\bibitem{kastnerbook2} R.E. Kastner,
{\emph{Adventures in Quantumland: Exploring Our Unseen Reality} (World Scientific, Singapore, 2019).}

\bibitem{TQ} T. Qureshi, ``Demystifying the delayed-choice quantum eraser,"
\href{https://doi.org/10.1088/1361-6404/ab923e}
{\emph{Eur. J. Phys.} \textbf{41}, 055403 (2020).}

\bibitem{Ma} X. Ma, J. Kofler, A. Zeilinger, ``Delayed-choice gedanken experiments and their realizations,"
\href{https://doi.org/10.1103/RevModPhys.88.015005}
{\emph{Rev. Mod. Phys.} \textbf{88}, 015005 (2016).}

\bibitem{vienna} X. Ma, J. Kofler, A. Qarry, N. Tetik, T. Scheidl, R. Ursin, S. Ramelow, T. Herbst, L. Ratschbacher, A. Fedrizzi, T. Jennewein, A. Zeilinger, ``Quantum erasure with causally disconnected choice,"
\href{https://doi.org/10.1073/pnas.1213201110}
{\emph{Proc. Natl. Acad. Sci. U.S.A.} \textbf{110}, 1221 (2013).}

\bibitem{mandel} A. G. Zajonc, L. Wang, X.Y Zou, L. Mandel, ``Quantum eraser,"
\href{https://doi.org/10.1038/353507b0}
{\emph{Nature} \textbf{353}, 507 (1991).}

\bibitem{chiao} P. G. Kwiat, A. Steinberg, R. Chiao, ``Observation of a quantum eraser: A revival of coherence in a two-photon interference experiment,"
\href{https://doi.org/10.1103/PhysRevA.45.7729}
{\emph{Phys. Rev. A} \textbf{45}, 7729 (1992).}

\bibitem{zeilinger} T. J. Herzog, P. G. Kwiat, H. Weinfurter, A. Zeilinger,
``Complementarity and the quantum eraser,"
\href{https://doi.org/10.1103/PhysRevLett.75.3034}
{\emph{Phys. Rev. Lett.} \textbf{75}, 3034 (1995).}

\bibitem{kim-shih} Y.-H. Kim, Rong Yu, Sergei P. Kulik, Y. Shih and
M. O. Scully, ``Delayed 'choice' quantum eraser,"
\href{https://doi.org/10.1103/PhysRevLett.84.1}
{\emph{Phys. Rev. Lett.} \textbf{84}, 1 (2000).}

\bibitem{walborn} S. P. Walborn, M. O. Terra Cunha, S. Pádua, C. H. Monken,
``Double-slit quantum eraser,"
\href{https://doi.org/10.1103/PhysRevA.65.033818}
{\emph{Phys. Rev. A} \textbf{65}, 033818 (2002).}

\bibitem{kim} H. Kim, J. Ko, and T. Kim, ``Quantum-eraser experiment with frequency-entangled photon pairs,"
\href{https://doi.org/10.1103/PhysRevA.67.054102}
{\emph{Phys. Rev. A} \textbf{67}, 054102 (2003).}

\bibitem{andersen} U. L. Andersen, O. Gl\"{o}ckl, S. Lorenz, G. Leuchs, R. Filip,
``Experimental demonstration of continuous variable quantum erasing,"
\href{https://doi.org/10.1103/PhysRevLett.93.100403}
{\emph{Phys. Rev. Lett.} \textbf{93}, 100403 (2004).}

\bibitem{scarcelli} G. Scarcelli, Y. Zhou, Y. Shih, ``Random delayed-choice quantum eraser via two-photon imaging,"
\href{https://doi.org/10.1140/epjd/e2007-00164-y}
{\emph{Eur. Phys. J. D} \textbf{44}, 167 (2007).}

\bibitem{neves} L. Neves, G. Lima, J. Aguirre, F.A. Torres-Ruiz, C. Saavedra,
A. Delgado, ``Control of quantum interference in the quantum eraser,"
\href{https://doi.org/10.1088/1367-2630/11/7/073035}
{\emph{New J. Phys.} \textbf{11}, 073035 (2009).}

\bibitem{schneider} M.B. Schneidera, I.A. LaPuma, ``A simple experiment for discussion of quantum interference and which-way measurement,"
\href{https://doi.org/10.1119/1.1450558}
{\emph{Am. J. Phys.} \textbf{70}, 266 (2002).}

\bibitem{bramon} A. Bramon, G. Garbarino, B.C. Hiesmayr, ``Quantum marking and quantum erasure for neutral kaons,"
\href{https://doi.org/10.1103/PhysRevLett.92.020405}
{\emph{Phys. Rev. Lett.} \textbf{92}, 020405 (2004).}

\bibitem{zini} T. Qureshi, Z. Rahman, ``Quantum eraser using a modified Stern-Gerlach setup,"
\href{https://doi.org/10.1143/PTP.127.71}
{\emph{Prog. Theor. Phys.} \textbf{127}, 71 (2012).}

\bibitem{barney} R.D. Barney, J-F. S. Van Huele ``Quantum coherence recovery through Stern–Gerlach erasure,"
\href{https://doi.org/10.1088/1402-4896/ab2d45}
{\emph{Phys. Scr.} \textbf{94}, 105105 (2019).}

\bibitem{chianello} M. Chianello, M. Tumminello, A. Vaglica, and G. Vetri,
``Quantum erasure within the optical Stern-Gerlach model,"
\href{https://doi.org/10.1103/PhysRevA.69.053403}
{\emph{Phys. Rev. A} \textbf{69}, 053403 (2004).}

\bibitem{3eraser} N.A. Shah, T. Qureshi, ``Quantum eraser for three-slit interference,"
\href{https://doi.org/10.1007/s12043-017-1479-8}
{\emph{Pramana J. Phys.} \textbf{89}, 80 (2017).}

\bibitem{terno} R. Ionicioiu, D.R. Terno, ``Proposal for a quantum delayed-choice experiment,"
\href{https://doi.org/10.1103/PhysRevLett.107.230406}
{\emph{Phys. Rev. Lett.} \textbf{107}, 230406 (2011).}

\bibitem{celeri} R. Auccaise, R.M. Serra, J.G. Filgueiras, R.S. Sarthour, I.S. Oliveira, L.C. Céleri, ``Experimental analysis of the quantum complementarity principle,"
\href{https://doi.org/10.1103/PhysRevA.85.032121}
{\emph{Phys. Rev. A} \textbf{85}, 032121 (2012).}

\bibitem{peruzzo} A. Peruzzo, P. Shadbolt, N. Brunner, S. Popescu, J.L. O’Brien, ``A quantum delayed-choice experiment,"
\href{https://doi.org/10.1126/science.1226719}
{\emph{Science} \textbf{338}, 634 (2012).}

\bibitem{kaiser} F. Kaiser, T. Coudreau, P. Milman, D.B. Ostrowsky, S. Tanzilli, ``Entanglement-enabled delayed-choice experiment,"
\href{https://doi.org/10.1126/science.1226755}
{\emph{Science} \textbf{338}, 637 (2012).}

\bibitem{qtwist} T. Qureshi, ``Quantum twist to complementarity: A duality relation,"
\href{https://doi.org/10.1093/ptep/ptt022}
{\emph{Prog. Theor. Exp. Phys.} \textbf{2013}, 041A01 (2013).}

\bibitem{guo} J-S. Tang, Y-L. Li, C-F. Li, G-C. Guo, ``Revisiting Bohr's principle of complementarity with a quantum device,"
\href{https://doi.org/10.1103/PhysRevA.88.014103}
{\emph{Phys. Rev. A} \textbf{88}, 014103 (2013).}

\bibitem{scarani} V. Scarani, A. Suarez, ``Introducing quantum mechanics: One-particle interferences"
\href{https://doi.org/10.1119/1.18938}
{\emph{Am. J. Phys.} \textbf{66}, 718 (1998). }

\bibitem{ferrari} C. Ferrari, B. Braunecker, ``Entanglement, which-way measurements, and a quantum erasure,"
\href{https://doi.org/10.1119/1.3369921}
{\emph{Am. J. Phys.} \textbf{78}, 792 (2010). }

\bibitem{hybrid} X. Ma, A. Qarry, J. Kofler, T. Jennewein, A. Zeilinger, ``Experimental Violation of a Bell Inequality with Two Different Degrees of Freedom of Entangled Particle Pairs,"
\href{https://doi.org/10.1103/PhysRevA.79.042101}
{\emph{Phys. Rev. A} \textbf{79}, 042101 (2009).}


\end{thebibliography}
\end{document}